\documentclass[12pt]{article}
\def\one{
\setlength{\unitlength}{0.45cm}
\begin{picture}(0.55,0.5)
\put(0,0){\line(1,0){0.4}}
\put(0,.4){\line(1,0){0.4}}
\multiput(0,0)(.4,0){2}{\line(0,1){.4}}
\end{picture}}
\def\twohor{
\setlength{\unitlength}{0.45cm}
\begin{picture}(1.1,0.5)
\put(0,0){\line(1,0){0.8}}
\put(0,.4){\line(1,0){0.8}}
\multiput(0,0)(.4,0){3}{\line(0,1){.4}}
\end{picture}}
\def\twover{
\setlength{\unitlength}{0.45cm}
\begin{picture}(0.55,0.5)
\put(0,0){\line(1,0){0.4}}
\put(0,.4){\line(1,0){0.4}}
\put(0,-.4){\line(1,0){0.4}}
\multiput(0,0)(.4,0){2}{\line(0,0){.4}}
\multiput(0,0)(.4,0){2}{\line(0,-1){.4}}
\end{picture}}
\def\threehor{
\setlength{\unitlength}{0.45cm}
\begin{picture}(1.5,0.5)
\put(0,0){\line(1,0){1.2}}
\put(0,.4){\line(1,0){1.2}}
\multiput(0,0)(.4,0){4}{\line(0,1){.4}}
\end{picture}}
\def\threever{
\setlength{\unitlength}{0.45cm}
\begin{picture}(0.6,0.5)
\put(0,0){\line(1,0){0.4}}
\put(0,.4){\line(1,0){0.4}}
\put(0,-.4){\line(1,0){0.4}}
\put(0,-.8){\line(1,0){0.4}}
\multiput(0,0)(.4,0){2}{\line(0,0){.4}}
\multiput(0,0)(.4,0){2}{\line(0,-1){.4}}
\multiput(0,0)(.4,0){2}{\line(0,-2){.8}}
\end{picture}}
\def\mixed{
\setlength{\unitlength}{0.45cm}
\begin{picture}(1,0.5)
\put(0,0){\line(1,0){0.8}}
\put(0,.4){\line(1,0){0.8}}
\put(0,-.4){\line(1,0){0.4}}
\multiput(0,0)(.4,0){3}{\line(0,1){.4}}
\multiput(0,0)(.4,0){2}{\line(0,-1){.4}}
\end{picture}}

\usepackage{epsf}
\usepackage{cite}
\def\one{
\setlength{\unitlength}{0.45cm}
\begin{picture}(0.55,0.5)
\put(0,0){\line(1,0){0.4}}
\put(0,.4){\line(1,0){0.4}}
\multiput(0,0)(.4,0){2}{\line(0,1){.4}}
\end{picture}}
\def\twohor{
\setlength{\unitlength}{0.45cm}
\begin{picture}(1.1,0.5)
\put(0,0){\line(1,0){0.8}}
\put(0,.4){\line(1,0){0.8}}
\multiput(0,0)(.4,0){3}{\line(0,1){.4}}
\end{picture}}
\def\twover{
\setlength{\unitlength}{0.45cm}
\begin{picture}(0.55,0.5)
\put(0,0){\line(1,0){0.4}}
\put(0,.4){\line(1,0){0.4}}
\put(0,-.4){\line(1,0){0.4}}
\multiput(0,0)(.4,0){2}{\line(0,0){.4}}
\multiput(0,0)(.4,0){2}{\line(0,-1){.4}}
\end{picture}}
\def\threehor{
\setlength{\unitlength}{0.45cm}
\begin{picture}(1.5,0.5)
\put(0,0){\line(1,0){1.2}}
\put(0,.4){\line(1,0){1.2}}
\multiput(0,0)(.4,0){4}{\line(0,1){.4}}
\end{picture}}
\def\threever{
\setlength{\unitlength}{0.45cm}
\begin{picture}(0.6,0.5)
\put(0,0){\line(1,0){0.4}}
\put(0,.4){\line(1,0){0.4}}
\put(0,-.4){\line(1,0){0.4}}
\put(0,-.8){\line(1,0){0.4}}
\multiput(0,0)(.4,0){2}{\line(0,0){.4}}
\multiput(0,0)(.4,0){2}{\line(0,-1){.4}}
\multiput(0,0)(.4,0){2}{\line(0,-2){.8}}
\end{picture}}
\def\mixed{
\setlength{\unitlength}{0.45cm}
\begin{picture}(1,0.5)
\put(0,0){\line(1,0){0.8}}
\put(0,.4){\line(1,0){0.8}}
\put(0,-.4){\line(1,0){0.4}}
\multiput(0,0)(.4,0){3}{\line(0,1){.4}}
\multiput(0,0)(.4,0){2}{\line(0,-1){.4}}
\end{picture}}

\catcode`@=11 \@addtoreset{equation}{section} \catcode`@=12

\def\p2{\frac{p}{2}}
\def\tp2{\frac{3p}{2}}

\begin{document}

\setlength \arraycolsep{2pt}

\begin{titlepage}
\vfill
\begin{center}
{\Large \bf Composite Representation Invariants and Unoriented   
Topological String Amplitudes}\\[1cm] 
Chandrima Paul\footnote{E-mail:chandrima@phy.iitb.ac.in}, 
Pravina Borhade \footnote{E-mail: borhadepravina@gmail.com}, 
P. Ramadevi\footnote{Email: ramadevi@phy.iitb.ac.in}\\
{\em Department of Physics, \\Indian Institute of Technology Bombay,\\
Mumbai 400 076, India\\[10pt]}
\end{center}
%\vspace{2cm}
\vfill
\begin{abstract}
Sinha and Vafa \cite {sinha} had conjectured that the $SO$ Chern-Simons 
gauge theory on $S^3$ must be dual to the closed $A$-model 
topological string on the orientifold of a resolved conifold. 
Though the Chern-Simons free energy could be rewritten in terms 
of the topological string amplitudes providing evidence
for the conjecture, we needed a novel idea in the 
context of Wilson loop observables to extract cross-cap 
$c=0,1,2$ topological amplitudes. Recent paper of Marino \cite{mar9} 
based on the work of Morton and Ryder\cite{mor} has clearly 
shown that the composite representation placed on the knots 
and links plays a crucial role to rewrite the topological 
string cross-cap $c=0$ amplitude. This enables extracting the 
unoriented cross-cap $c=2$ topological  amplitude. In this paper, 
we have explicitly worked out the composite invariants 
for some framed knots and links carrying 
composite representations in $U(N)$ Chern-Simons theory. 
We have verified generalised Rudolph's theorem,
which relates composite  invariants to the invariants in 
$SO(N)$ Chern-Simons theory, and also verified Marino's conjectures
on the integrality properties of the topological
string amplitudes. For some framed knots and links, we have 
tabulated the BPS integer invariants for cross-cap $c=0$ and 
$c=2$ giving the open-string topological amplitude 
on the orientifold of the resolved conifold.
\end{abstract}
\vfill
\end{titlepage}
\section{Introduction}
We have seen interesting developments in the open string and 
closed string dualities during the last 12 years starting
from the celebrated work of Maldacena \cite{malda}. 
Gopakumar and Vafa \cite {gv1,gv2,gv3} conjectured  open-closed 
duality in the topological string context. 
Gopakumar-Vafa conjecture states that the $A$-model
open topological string theory on the deformed conifold,
equivalent to  the Chern-Simons gauge theory on $S^3$ \cite {wittencs}, 
is dual to the closed string theory on a resolved conifold. 

In ref. \cite {gv1}, it was shown that the free-energy expansion 
of $U(N)$ Chern-Simons field theory on $S^3$ at large $N$  resembles 
$A$-model topological string theory amplitudes on the resolved conifold.
This provided an evidence for the conjecture. Another piece of
evidence at the level of observables was shown by Ooguri and 
Vafa \cite{ov} for the simplest Wilson loop observable (simple 
circle also called unknot) in Chern-Simons theory on $S^3$. 
In particular, Ooguri-Vafa considered the expectation value of a 
scalar operator ${\cal Z}_{\cal H}(v)$ in the topological 
string theory corresponding to the simple circle in submanifold 
$S^3$ of the deformed conifold and showed its  form in the resolved 
conifold background. From these results for unknot,
Ooguri-Vafa conjectured on the form for  
${\cal Z}_{\cal H}(v)$ for any knot or link in $S^3$.
For completeness and simplicity, we briefly present the form for knots:  
\begin{eqnarray}
{\cal F}_{\cal H}(v)&=&\ln {\cal Z}_{\cal H}(v)=\ln\{
\sum_R {\cal H}_R[{\cal K}] s_R(v)\} 
~=\sum_{R,d} f_R(q^d,\lambda^d) s_R(v^d)~\\
{\rm where}~f_R(q,\lambda)&=& {1 \over (q^{1/2}-q^{-1/2})}
\sum_{Q,s}N_{R,Q,s} \lambda^Q q^s \label {unrfm}
\end{eqnarray}
Here ${\cal H}_R({\cal K})$ are the $U(N)$ Chern-Simons invariants
for a knot ${\cal K}$ in $S^3$ carrying representation $R$ and $s_R(v)$
are the Schur polynomials in variable $v$ which represent
$U(N)$ holonomy of the knot ${\cal K}$ in the Lagragian submanifold
${\cal N}$ which intersects $S^3$ along the knot. ${\cal F}_{\cal H}(v)$
denotes the free-energy of the topological open-string partition function
on the resolved conifold and $f_R(q,\lambda)$ are the
$U(N)$ reformulated invariants. The conjecture states that the 
reformulated invariant must have the form (\ref {unrfm}) where 
$N_{R,Q,s}$ are integer coefficients.

Labastida-Marino \cite{lm} used group-theoretic 
techniques to rewrite the expectation value of the topological operators 
in terms of link invariants in $U(N)$ Chern-Simons field theory
on $S^3$.  This group theoretic approach enabled verification of 
Ooguri-Vafa conjecture for many non-trivial knots\cite{lm, taps, laba, mari1}. 
Conversely, the Ooguri-Vafa conjecture led to a reformulation of 
Chern-Simons field theory invariants for knots and links giving new
polynomial invariants(\ref {unrfm}). The integer coefficients of these
new polynomial invariants have topological meaning accounting
for BPS states in the string theory. The challenge still  
remains in obtaining such integers for non-trivial knots
and links within topological string theory.

Another challenging question is to attempt 
similar duality conjectures between Chern-Simons gauge theories 
on three-manifolds other than $S^3$ and closed string theories. 
Invoking Gopakumar-Vafa conjecture and Ooguri-Vafa conjecture,  
it was possible to explicitly write the $U(N)$ Chern-Simons 
free-energy expansion at large $N$ as a closed string 
theoretic expansion\cite{ramprav}. Surprisingly, the  expansion 
resembled partition function of a closed string theory on a 
Calabi-Yau background with one kahler parameter. 
Unfortunately,  the Chern-Simons free-energy expansion 
for other three-manifolds are not equivalent to the `t Hooft 
large $N$ perturbative expansion around a classical solution \cite {thoo}.
In order to predict new duality conjectures, we need to extract 
the perturbative expansion around a classical solution from the 
free-energy.

For orbifolds of $S^3$, which gives Lens space ${\cal L}[p,1] \equiv S^3/Z_p$,
it is believed that the Chern-Simons theory is dual to the
$A$-model closed string theory on $A_{p-1}$ 
fibred over $P^1$ Calabi-Yau background.  It was Marino \cite {marin} who
showed that the perturbative Chern-Simons theory on 
Lens space ${\cal L}[p,1]$ can be given a  matrix model description. 
Also, hermitian matrix model description of $B$-model topological
strings \cite {dijk} was shown to be equivalent to Marino's matrix model
using mirror symmetry\cite{akmv}. It is still a challenging open problem 
to look for dual closed string description corresponding to $U(N)$ Chern-Simons
theory on other three-manifolds.

The extension of these duality conjectures for other gauge 
groups like $SO(N)$ and $Sp(N)$ have also been studied.
In particular, the free-energy expansion $F^{(SO)}_{(CS)}[S^3]$ of
the Chern-Simons theory on $S^3$ based on $SO$ gauge group was 
shown to be dual to $A$-model closed string theory on a orientifold 
of the resolved conifold background \cite {sinha}. In particular, 
the string partition function $Z$ for these orientifolding action 
must have two contributions: 
\begin{equation}
F^{(SO)}_{(CS)}[S^3]~=Z~=~{1 \over 2}Z^{or} + Z^{(unor)} \label {sinh}
\end{equation}
where $Z^{(or)}$ is the untwisted contribution and $Z^{(unor)}$
is the twisted sector contribution. The untwisted contribution
exactly matches the $U(N)$ Chern-Simons free energy on $S^3$. 
Using the topological vertex as a tool, Bouchard et al \cite {vinc1,vinc2} 
have determined unoriented closed string amplitude and unoriented 
open topological string amplitudes for a few orientifold toric geometry 
with or without $D$-branes. 

In Ref.\cite {prav}, the generalisation of Ooguri-Vafa conjecture 
for observables involving $SO(N)$ holonomy, different 
from the works of Bouchard et al \cite{vinc1,vinc2}, was studied.
Similar to the $U(N)$ result (\ref {unrfm}), the coefficients of 
$SO(N)$ reformulated invariants are indeed integers.  

Following Sinha-Vafa conjecture \cite{sinha}, the expectation value of the 
topological string operator (observables) $Z_{\cal G}(v)$ 
where ${\cal G}$ represents $SO(N)$ knot invariants in 
Chern-Simons theory on $S^3$ and $v$ represents the $SO$ holonomy on
the submanifold ${\cal N}$ intersecting $S^3$ along a knot. It is expected
that the free-energy of the open-string partition function on the orientifold
of the resolved conifold must also satisfy a relation 
similar to eqn.(\ref{sinh}): 
\begin{equation}
{\cal F}_{\cal G}(v)= \ln Z_{\cal G}(v)= {1 \over 2} {\cal F}^{(or)}_{\cal R}(v)
+{\cal F}^{(unor)}(v)~. \label {orie1} 
\end{equation}
where ${\cal F}^{(or)}_{\cal R}(v)$ is the oriented or
untwisted sector contribution (also called cross-cap $c=0$)
and the twisted sector term ${\cal F}^{(unor)}(v)$ will have both 
cross-cap $c=1$ and $c=2$ contributions to the open topological
string amplitudes.
It was not clear \cite{vinc1,vinc2} as to how to obtain 
${\cal F}^{(or)}_{\cal R}(v)$ in the orientifold theory 
using $U(N)$ Chern-Simons knot invariants.
As a result, it was not possible to distinguish the topological
amplitudes of cross-cap $c=0$ from 
$c=2$ contribution. However using parity argument
in variable $\sqrt{\lambda}$, the cross-cap $c=1$ topological
amplitudes contribution could be obtained\cite{vinc1,vinc2,prav}. 

From the orientifolding action, Marino \cite{mar9} has indicated 
that there must be a $U(N)$ composite
representation $(R,S)$ placed on the knot in $S^3$ and the
oriented contribution must be rewritable as: 
\begin{equation}
{\cal F}^{(or)}_{\cal R}(v)=\sum_{R,S}{\cal H}_{(R,S)}[{\cal K}] s_R(v) s_S(v)
=\sum_R {\cal R}_R[{\cal K}] s_R(v) \label {orient}
\end{equation}
where $s_R(v)$ and $s_S(v)$ are the Schur polynomials 
corresponding to the $U(N)$ holonomy in two Lagrangian
submanifolds ${\cal N}_{\epsilon}$ and 
${\cal N}_{-\epsilon}$ related by the orientifolding action.
Here $\epsilon$ denotes the deformation parameter of the 
deformed conifold. The oriented invariant ${\cal R}_R[{\cal K}]$
can be obtained from composite invariants ${\cal H}_{(R,S)}[{\cal K}]$ 
using the properties of the Schur polynomials. Though we have so far 
discussed for knots, it is straightforward to generalise 
these arguments for any $r$-component link $L$.

In this paper, we explicitly evaluate the 
composite invariants ${\cal H}_{(R_1,S_1),(R_2,S_2),\ldots (R_r,S_r)}[L]$, 
in $U(N)$ Chern-Simons gauge theory for many framed knots and links $L$
made of $r$ component knots ${\cal K}_{\alpha}$'s carrying
composite representations $(R_{\alpha},S_{\alpha})$ 
using the tools\cite{rama}. These composite invariants
are polynomials in two variables $q,\lambda$. 
We find that the framing factor for the component knots of the 
links carrying composite representation requires a 
slightly modified choice of the $U(1)$ charge so that the 
composite invariants are polynomials in variables
$q$ and $\lambda$.  

Comparing these invariants with $SO(N)$ Chern-Simons invariants 
${\cal G}_{R_1,R_2,\ldots R_r}[L]$\cite{prav} for link
$L$ whose components carry representations $R_{\alpha}$'s which are 
also polynomials in two variables $(q,\lambda)$, 
we have verified the generalised Rudolph's theorem\cite{mor,Rudo}:
\begin{equation}
{1 \over 2} \left[ {\cal H}_{(R,R)}[{\cal K}]+
\{{\cal G}_R[{\cal K}]\}^2 \right]=
f(q)\sum_{n,p} a_{n,p} \lambda^{n \over 2} q^p~, \label {rud}
\end{equation}
for many framed knots ${\cal K}$ carrying $R=\twover, \twohor, \one$. 
Here $f(q)$ is a function of $q$, $a_{n,p}$ are integers. In fact, the 
above relation between $U(N)$ composite invariants and 
$SO(N)$ invariants appears naturally from the integrality properties 
of the topological string amplitudes in the orientifold geometry \cite{mar9}.
Using these composite representation invariants, we  
verified the integrality conjectures of Marino\cite{mar9}
for  framed knots and framed two-component links.  
While submitting this paper, we came across a recent paper\cite {stev}
where Marino's conjectures have been verified for standard framing 
torus knots and torus links which is a special case of our results.

The organisation of the paper is as follows. In
section 2, we present composite framed knot and 
framed two-component link invariants in $U(N)$ Chern-Simons theory.  
In section 3, we briefly review Marino's conjectures
on the reformulated invariants of the framed links 
in the orientifold resolved conifold.
In section 4, we verify Marino's conjectures
and tabulate the $c=0$,$c=2$ BPS integer coefficients
for few examples. In the concluding section, we summarize the results
obtained.  In appendix A, we present $U(N)$ composite invariants  for
some framed knots and framed two-component links for some representations.
 
\section{Chern-Simons Gauge theory and Composite Link invariants}
Chern-Simons gauge theory on $S^3$ based on the gauge group $G$ 
is described by the following action:
\begin{equation}
S = {k \over 4 \pi} \int_{S^3} Tr\left (A \wedge dA + {2 \over 3} A \wedge
A \wedge A \right)  
\end{equation}
where $A$ is a gauge connection for compact semi-simple gauge group $G$ and 
$k$ is the coupling constant. The observables in this theory are 
Wilson loop operators: 
\begin{equation}
W_{R_1,R_2, \ldots R_r}[L]~=~ \prod_{\alpha=1}^rTr_{R_{\alpha}} 
U [{\cal K}_{\alpha}]~,
\end{equation}
where $U[{\cal K}_{\alpha}]=P\left[\exp \oint_{{\cal K}_{\alpha}} A\right]$
denotes the holonomy of the gauge field $A$ around the 
component knot ${\cal K}_{\alpha}$
of a $r$-component link $L$  carrying 
representation $R_{\alpha}$. The expectation value of these Wilson loop 
operators are the link invariants:
\begin{equation}
\langle W_{R_1,R_2, \ldots R_r}[L] \rangle(q,\lambda)= {\int[{\cal D}A]e^{iS}
 W_{R_1,R_2,\ldots,R_r}[L] \over \int[{\cal D} A]e^{iS}}~, 
 \label {linki} 
\end{equation}
These link invariants are polynomials in two variables 
\begin{equation}
q=\exp\left({2 \pi i \over k+C_v}\right)~,~ \lambda = q^{N+a}~,
\end{equation}
where $C_v$ is the dual coxeter number of the gauge group $G$
\begin{displaymath}
\begin{array} {lcl}
C_v = \left\{ \begin{array}{ll}
N & {\rm for}~ G=SU(N)\\
N-2 & {\rm for} ~G=SO(N)\\
\end{array} \right. & {\rm and} &
a = \left\{ \begin{array}{ll}
0 & {\rm for}~ G=SU(N)\\
-1 & {\rm for} ~G=SO(N)\\
\end{array} \right.
\end{array}
\end{displaymath}
These link invariants can be computed using the following two 
inputs\cite{rama}:\\
(i) Any link can be drawn as a closure or plat of braids,~\\
(ii) The connection between Chern-Simons theory and the Wess-Zumino 
conformal field theory.\\
We now define some quantities which will be useful later. The quantum
dimension of a representation $R$ with
highest weight $\Lambda_R$ is given by
\begin{equation}
dim_q R=\Pi_{\alpha>0}\frac{[\alpha\cdot (\rho+\Lambda_R)]
}{[\alpha\cdot\rho]}~, \label {quan}
\end{equation}
where $\alpha$'s are the positive roots and $\rho$ is the Weyl vector
equal to the sum of the fundamental weights of the group $G$.
The square bracket refers to the quantum number defined by
\begin{equation}
[x]={\left(q^{x/2}-q^{-x/2} \right) \over \left(q^{1/2}- q^{-1/2} \right)}~.
\label{qno}
\end{equation}
The $SU(N)$ quadratic Casimir for representation $R$ is given by
\begin{equation}
C_R = -\frac{\ell^2}{2N}+\kappa_R =  -\frac{\ell^2}{2N}+
\frac{1}{2} \left ((N+a) \ell + \ell + \sum_i (l_i^2- 2i l_i) \right)~.
\label {casi}
\end{equation}
Our interest is to obtain invariants of framed knots and framed links
carrying representation $R_c \equiv (R,S)$ called composite 
representation in $U(N)$ Chern-Simons gauge theory so that
Marino's conjectures on the topological amplitudes
in the orientifold of resolved conifold geometry can be
verified.
\subsection{Composite Invariants in $U(N)$ Chern-Simons Gauge Theory}
The composite representation, $R_c \equiv (R,S)$ labelled by a 
pair of Young diagram is defined as \cite{mar9,compo,mor1,mor2}
\begin{equation}
R_c \equiv (R,S)= {\sum_{U,V,W}} {{(-1)}^{l(U)}}N^R_{U V} N^{S}_{{{U}^T}W} 
(V \times \bar W)~, \label {composite}
\end{equation}
where $U,V,W$ are the representations of the group U(N) , $\ell(U)$ denotes
the number of boxes in the Young diagram corresponding to $U$ and $N$ is
the Littlewood-Richardson coefficient for multiplication of the Young diagrams.

If we take the simplest defining representation 
for $R=\one$ and $S=\one$, then the composite representation $R_c=
(\one,\one)$ derived from eqn. (\ref{composite}) will be the adjoint 
representation of $U(N)$. In terms of fundamental
weights, the highest weight of $R_c$ is $\Lambda^{(1)}+\Lambda^{(N-1)}$. 
Using the above eqn.(\ref {composite}), one can obtain the 
$SU(N)$ representation for any composite representation $(R,S)$ and the 
corresponding highest weight will be $\Lambda_R + \Lambda_{\bar S}$ 
where $\Lambda_R$ and $\Lambda_{\bar S}$
are the highest weights of representation $R$ and conjugate representation
$\bar S$ respectively.

We will now explicitly evaluate the polynomials for various knots
and links carrying the composite representation $(R,S)$ in $U(N)$ Chern-Simons
theory. For the simplest circle called unknot $U_p$ with an arbitrary framing 
$p$, the composite invariant will be framing factor multiplying
the quantum dimension of the composite representation $(R,S)$: 
\begin{equation}
{{\cal H}_{(R,S)}}[U_p] = (-1)^{\ell p} q^{p \{n_{(R,S)}\}^2\over 2}
q^{p C_{(R,S)}}dim_q (R,S)~. \label {frame}
\end{equation}
where $\ell$ is the total number of boxes in the Young diagram
for composite representation $(R,S)$, $C_{(R,S)}$ denotes the $SU(N)$ 
quadratic casimir (\ref {casi})  and 
$n_{(R,S)}$ represents the $U(1)$ charge for the composite representation
$(R,S)$. Looking at the definition of the composite representation highest
weight, we propose that the $U(1)$ charge $n_{(R,S)}$ must be the
difference of $U(1)$ charges $n_R$ and $n_S$ of representation $R$ and
$S$:
\begin{equation}
n_{(R,S)}=|n_R-n_S|~.\label {framchar}
\end{equation}
Earlier the $U(1)$ charges for $U(N)$ representations were 
chosen\cite{mari1,ramprav} such the $U(N)$ invariants are 
polynomials in two variables $q,\lambda$ \cite{mari1,ramprav}. 
For representation $R$ with $\ell(R)$ number of
boxes in the Young diagram representation, the $U(1)$ charge $n_R$ is
\begin{equation}
n_R={\ell(R) \over \sqrt N}~. \label{u1char}
\end{equation}  
Substituting the $U(1)$ charge (\ref{u1char}) in eqn.(\ref {framchar}),
the unknot invariant (\ref {frame}) simplies to
\begin{equation}
{{\cal H}_{(R,S)}}[U_p] = (-1)^{\ell p} q^{\kappa_R+\kappa_S} dim_q (R,S)~.
\label {frame1}
\end{equation}
Our choice for the  composite representation $U(1)$ charge (\ref{framchar})
results in the simplied form for the framing factor. For knots
carrying composite representation $(R,S)$, the framing
factor in eqn.(\ref {frame1}) involves only the sum of 
$\kappa_R$ and $\kappa_S$ (\ref {casi}).

Now, we can write the $U(N)$ framed knot invariants for torus knots 
${\cal K}_{2m+1}^{(p)}$ of the type $(2, 2m+1)$ with framing $p$ as follows:
\begin{equation}
{{\cal H}_{(R,S)}}[{\cal K}_{2m+1}^{(p)}](q,\lambda)= (-1)^{\ell p} q^{p (\kappa_R+
\kappa_S)} \sum_{R_t}  
dim_q R_t (\lambda_t)^{2m+1}~, \label {knotcom}
\end{equation} 
where $R_t \in (R,S) \otimes (R,S)$ and $\lambda_t$ 
is the braiding eigenvalue in standard framing ($p=0$) for
parallely oriented strands:
\begin{equation}
\lambda_t= \epsilon_{R_t} q^{2C_{(R,S)}-C_{R_t}/2}~ ,\label{egval}
\end{equation}
where $\epsilon_{R_t}= \pm 1$ depending upon whether the representation $R_t$
appears symmetrically or antisymmetrically with respect to the tensor
product $(R,S) \otimes (R,S)$ in the $U(N)_k$ Wess-Zumino Witten model.
Unlike the totally symmetric or totally antisymmetric representations,
the tensor product of composite representations does give multiplicities
and hence determining $\epsilon_{R_t}$ is non-trivial.

We have fixed the sign of $\epsilon_{R_t}$ by equating the invariants
of two knots which are equivalent. For example, unknot $U_0$ and the 
torus knot ${\cal K}_1^{(0)}$ are equivalent. Taking the difference
of these two knot polynomials and equating the coefficients
of every power of $q$ to zero, we obtain the signs of $\epsilon_{R_t}$ 
uniquely. In appendix A, we have explicitly given all the 
irreducible representations $R_t$ and the signs $\epsilon_{R_t}$ for 
some composite representations so that the composite invariants can 
be computed. This will be very useful for verifying Rudolph theorem and 
Marino's conjectures. We can also check Marino's conjecture for 
composite invariant for connected sum of two knots ${\cal K}_1$ and 
${\cal K}_2$ defined as:
\begin{equation}
{\cal H}_{(R,S)}[{\cal K}_1\# {\cal K}_2]=\left( 
{\cal H}_{(R,S)}[{\cal K}_1] {\cal H}_{(R,S)}[{\cal K}_2]\right)/ 
{\cal H}_{(R,S)}[U_0]~.
\end{equation}
The $U(N)$ invariants for framed torus links of the type $(2,2m)$
can also be written. For example, the $U(N)$ invariant for a 
Hopf link of type $(2,2)$ with linking number $-1$ and framing numbers 
$p_1$ and $p_2$ on the component knots carrying representations 
$(R_1,S_1)$ and $(R_2,S_2)$ will be 
\begin{eqnarray}
{\cal H}_{(R_1,S_1),(R_2,S_2)}[H](q,\lambda)&=& (-1)^{\ell_1 p_1 + \ell_2 p_2} 
q^{p_1 (\kappa_{R_1}+\kappa_{S_1})+p_2 (\kappa_{R_2}+\kappa_{S_2})} \times\nonumber\\
~&~&q^{\ell k n_{(R_1,S_1)}n_{(R_2,S_2)}}\sum_{R_t} 
dim_q R_t q^{C_{(R_1,S_1)} + C_{(R_2,S_2)}- C_{R_t}}~, \label {link}
\end{eqnarray}
where $\ell k=-1$ is the linking number between the two-components
and $R_t \in (R_1,S_1)\otimes(R_2,S_2)$.
We now explicitly evaluate the knot polynomials carrying the composite
representation $(\one,\one)$ in $U(N)$ Chern-Simons theory, for the knots
upto five crossings. For the simplest composite representation 
$(\one,\one)$, which we denote by  $\rho_0$, the highest weight is 
$\Lambda^{(N-1)}+\Lambda^{(1)}~.$ 
The $p$-frame unknot invariant for this representation is
\begin{equation}
{\cal H}_{(\one,\one)} [U_p]= (-1)^{\ell p}\lambda^{p}(dim_q \rho_0)=
(-1)^{Np}\lambda^{p}[N-1][N+1]~,
\end{equation}
where rewriting the quantum numbers (\ref {qno}) will give the $p$-framed
unknot invariant in variables $q,\lambda=q^N$.
The highest weights for all the representations $R_t$'s obtained from 
$\rho_0 \otimes \rho_0$ and their corresponding
quantum dimensions(\ref{quan}) with the braiding eigenvalues 
(\ref {egval}) are tabulated: 

\noindent
\begin{tabular}{|l|l|l|l|}\hline
$R_t$&Highest weight&$dim_qR_t$& $\lambda_t$ \\
\hline
~&~&~&\\ 
$R_1$ & ${\Lambda^{(N-2)}}+2\Lambda^{(1)}$& 
$\frac{[N-1][N-2][N+1][N+2]}{[2][2]}$& $-\lambda~$\\ 
~&~&~&\\ 
$R_2$& $2\Lambda^{(N-1)}+{\Lambda^{(2)}}$&
$\frac{[N-1][N-2][N+1][N+2]}{[2][2]}$ & $-\lambda~$\\
~&~&~&\\ 
$R_3$ & ${\Lambda^{(N-2)}}+\Lambda^{(2)}$&
$\frac{[N]^2[N-3][N+1]}{[2][2]}$ & $q\lambda~$\\
~&~&~&\\ 
%~&~&~&\\ 
\hline
\end{tabular} \hspace{.001cm}
\begin{tabular}{|l|l|l|l|}\hline
$R_t$&Highest weight&$dim_q R_t$& $\lambda_t$ \\
\hline 
$R_4$ & ${\Lambda^{(N-1)}}+\Lambda^{(1)}$&
$[N-1][N+1];$ & $\lambda^{3/2}$ \\
$R_5$ &$2 \Lambda^{(N)}$&
$ 1$  & $\lambda^2~$ \\
~&~&~&\\ 
$R_6$& $2\Lambda^{(N-1)}+2{\Lambda^{(1)}}$&
$\frac{[N]^2[N+3][N-1]}{[2][2]}$ & $q^{-1} \lambda~$ \\
~&~&~&\\ 
$R_7$ & ${\Lambda^{(N-1)}}+\Lambda^{(1)}$&
$[N-1][N+1]$ & $-\lambda^{3/2}$\\
~&~&~&\\ 
\hline 
\end{tabular}
\vskip.2cm
Substituting the tabulated data in eqn.(\ref{knotcom}), the knot 
invariants for the framed knot ${\cal K}_{2m+1}^{(p)}$  
carrying representation $\rho_0 = (\one, \one)$ can be computed.
We have presented the tensor products
for other composite representations in the 
appendix A. This data will be very useful to directly compute the composite
invariants of framed knots (\ref {knotcom}) and framed links (\ref {link}).
These results are new and they are very essential to verify 
generalised Rudolph theorem (\ref {rud}) for many knots and links. 
The composite invariants also play a crucial role in verifying
Marino's conjecture and obtaining the topological 
string amplitudes corresponding to cross-caps
$c=0,1~ {\rm and}~2$. 

Using these $U(N)$ composite invariants
and the $SO(N)$ invariants in appendix A of Ref.\cite{prav}, we have verified
generalised Rudolph's theorem (\ref {rud}). In the next section we 
will recaptitulate the essential ideas of Marino's proposal 
for obtaining the cross-cap $c=0$ and $c=2$ topological amplitudes. 
\section{ Reformulated Link Invariants}
We will now review the conjectures proposed by Marino {\cite{mar9}}
for the reformulated $SO(N)$ invariants of knots and links.
Particularly, we have to get the untwisted sector (oriented) contribution
(\ref {orie1})to the open topological string amplitudes on the orientifold
of the resolved conifold geometry. 

Using the properties satisfied by Schur polynomials,
eqn.(\ref {orient}) implies that the oriented invariants
${\cal R}_{R_1,\ldots,R_r}[L]$ of the link $L$ whose
components ${\cal K}_1,\ldots,{\cal K}_r$ are colored by
representations $R_1,\ldots,R_r$ is given by
\begin{equation} 
{\cal R}_{R_1,\ldots,R_r} [L] = \sum_{S_1,T_1,\ldots,S_r,T_r} 
\prod_{\alpha=1}^r N^{R_\alpha}_{S_\alpha,T_\alpha} 
{\cal H}_{(S_1,T_1),\ldots,(S_r,T_r)}[L]~, \label {orie2}
\end{equation}
where $N^{R_\alpha}_{S_\alpha,T_\alpha}$ are the Littlewood-Richardson 
coefficients and ${\cal H}_{(S_1,T_1),\ldots,(S_r,T_r)}[L]$ are composite
invariants in $U(N)$ Chern-Simons gauge theory
of the link whose components carry the composite
representations $(S_1,T_1),\ldots,(S_r,T_r)$ of $U(N)$.
The generating functional giving the oriented contribution to the 
open topological string partition function (\ref {orie1}) is defined as
\begin{equation}
{\cal Z}_{\cal R}(v_1,\ldots,v_r) = \sum_{R_1,\ldots,R_r} 
{\cal R}_{R_1,\ldots,R_r} [L]\,
\prod_{\alpha=1}^r s_{R_{\alpha}}(v_{\alpha}); \,\,
{\cal F}_{\cal R}(v_1,\ldots,v_r) = {\rm log} \, {\cal Z}_{\cal R}(v_1,\ldots,v_r)~,
\end{equation} 
where $s_R(v)$ are the Schur polynomials. Also the generating 
functionals for those involving $SO(N)$ Chern-Simons invariants , 
${\cal G}_{R_1,\ldots,R_r}$, of a link $L$ are defined as
\begin{equation} 
{\cal Z}_{\cal G}(v_1,\ldots,v_r) = \sum_{R_1,\ldots,R_r} 
{\cal G}_{R_1,\ldots,R_r} [L] \, 
\prod_{\alpha=1}^r s_{R_{\alpha}}(v_{\alpha}); \,\,
{\cal F}_{\cal G}(v_1,\ldots,v_r) = {\rm log} \, {\cal Z}_{\cal G}(v_1,\ldots,v_r)~.
\end{equation} 
Marino \cite{mar9} has conjectured a specific form for
these generating functionals:
\begin{equation} 
{\cal F}_{\cal R}(v_1,\ldots,v_r) = \sum_{d=1}^{\infty} \sum_{R_1,\ldots,R_r}
h_{R_1,\ldots,R_r} (q^d,\lambda^d) \, \prod_{\alpha=1}^r s_{R_{\alpha}}
(v_{\alpha}^d)~, \label {con1a}
\end{equation}
and
\begin{equation}
{\cal F}_{\cal G}(v_1,\ldots,v_r) - \frac{1}{2} {\cal F}_{\cal R}(v_1,\ldots,v_r) = \sum_{d\,odd} \sum_{R_1,\ldots,R_r}
g_{R_1,\ldots,R_r} (q^d,\lambda^d) \, \prod_{\alpha=1}^r s_{R_{\alpha}}
(v_{\alpha}^d)~, \label {con2a}
\end{equation}
where $h_{R_1,\ldots,R_r} (q,\lambda)$ and $g_{R_1,\ldots,R_r} (q,\lambda)$
are the reformulated polynomial invariants involving the $U(N)$
and $SO(N)$ Chern-Simons link invariants respectively. The reformulated
invariants are polynomials in $q$ and $\lambda$ and conjectured to 
obey the following form
\begin{equation}
h_{R_1,\ldots,R_r} (q,\lambda) \,{\rm or}\, g_{R_1,\ldots,R_r} (q,\lambda)=
\sum_{Q,s} \frac{1}{q^{1/2}-q^{-1/2}} \tilde{N}_{R_1,\ldots,R_r,Q,s}
q^s\,\lambda^Q\,,\label{horg}
\end{equation}
where $\tilde{N}_{R_1,\ldots,R_r,Q,s}$ are integers.
Though we know that the reformulated invariants
$f_R(q,\lambda)$ obtained from $U(N)$ invariants ${\cal H}_R[L]$ satisfies
the conjecture (\ref {unrfm}), it is not at all obvious that 
the reformulated invariant 
$h_{R_1,\ldots,R_r} (q,\lambda)$ corresponding to 
the oriented invariants (\ref {orie2}) involving linear
combination of $U(N)$ composite invariants must obey a similar
conjectured form (\ref {horg}). We check few examples in section 4 to 
verify Marino's conjecture on the oriented reformulated invariants.
These reformulated invariants are further refined using the following 
equations, in order to reveal the BPS structure
\begin{eqnarray} 
h_{R_1,\ldots,R_r}(q,\lambda) &=& \sum_{S_1,\ldots,S_r}
 M_{R_1,\ldots,R_r;S_1,\ldots,S_r} \, \hat{h}_{S_1,\ldots,S_r}(q,\lambda),\label{reh}\\
g_{R_1,\ldots,R_r}(q,\lambda) &=& \sum_{S_1,\ldots,S_r}
 M_{R_1,\ldots,R_r;S_1,\ldots,S_r} \, \hat{g}_{S_1,\ldots,S_r}(q,\lambda)~,\label{reg}
\end{eqnarray} 
where
\begin{equation}
M_{R_1, \ldots R_r;S_1,\ldots S_r} = \sum_{T_1, \ldots
T_r} \prod_{\alpha=1}^r C_{R_{\alpha} S_{\alpha} T_{\alpha}}
S_{T_{\alpha}}(q)~,
\end{equation}
$R_{\alpha},S_{\alpha}, T_{\alpha}$ are
representations of the symmetric group $S_{\ell_{\alpha}}$ which can be
labelled by a Young-Tableau with a total of $\ell_{\alpha}$ boxes
and $C_{RST}$ are the Clebsch-Gordan coefficients of the symmetric
group. $S_R(q)$ is non-zero only for the hook representations. For a hook
representation having $\ell -d$ boxes in the first row of Young tableau
with total $\ell$ boxes, $S_R(q)=(-1)^d q^{-(\ell-1)/2+d}$. Marino
\cite{mar9} has conjectured that the refined reformulated invariants 
$\hat{h}_{R_1,\ldots,R_r} (q,\lambda)$ 
and $\hat{g}_{R_1,\ldots,R_r} (q,\lambda)$ should have the following structure:
\begin{eqnarray}
\hat{h}_{R_1,\ldots,R_r} (q,\lambda) &=& z^{r-2} \sum_{g\geq 0} \sum_{Q}
N^{c=0}_{R_1,\ldots,R_r,g,Q} z^{2g} \lambda^Q~, \label{hath} \\
\hat{g}_{R_1,\ldots,R_r} (q,\lambda) &=& z^{r-1} \sum_{g\geq 0} \sum_{Q}
\left( N^{c=1}_{R_1,\ldots,R_r,g,Q} z^{2g} \lambda^Q 
+ N^{c=2}_{R_1,\ldots,R_r,g,Q} z^{2g+1} \lambda^Q\right)~,\label{hatg}
\end{eqnarray}
where $N^{c=0}_{R_1,\ldots,R_r,g,Q}$, $N^{c=1}_{R_1,\ldots,R_r,g,Q}$ and
$N^{c=2}_{R_1,\ldots,R_r,g,Q}$ are the BPS invariants corresponding to
cross-caps $c=0,1 \,{\rm and}\, 2$ respectively and the variable
$z=q^{1/2}-q^{-1/2}$.

In the next section, we obtain the reformulated invariants and
obtain the BPS integers coefficients for framed knots and framed
two-component links.

\section{Verification of Marino's conjectures}
The composite polynomials which we computed with our proposed  choice
of framing factor obeys the conjectures of Marino. We will briefly
present some examples in this section.
\subsection{Computation of Oriented Invariants $h_{R_1,\ldots,R_r}(q,\lambda)$
and BPS invariants $N^{c=0}_{R_1, \ldots R_r,g,Q}$}
In this subsection, we list the reformulated oriented invariants and 
the corresponding BPS invariants for simple framed knots 
to verify the conjecture (\ref {horg}).

Using the invariants for knots carrying composite representations as
detailed in section 2 and appendix A, it is straightforward to obtain the 
reformulated oriented invariants.
We have checked that these invariants for the torus knots and
two-component links obey Marino's conjectured form. For completeness,
we shall present some of the oriented reformulated invariants (\ref {horg}) 
for the unknot with framing $p$ ($U_p$):  
\begin{eqnarray} 
h_{\twohor}[U_p]&=& \frac{-\lambda^{p-1}}{{\left( -1 + q \right) }^2\,\left( 1 + q \right) } 
\left( \lambda  - 2\,q^{1 + p}\,\left( -1 + \lambda  \right) \,\left( -1 + q\,\lambda  \right)  + 
  {\left( -1 \right) }^p\,\left( -1 + q \right) \,q\,\left( -1 + {\lambda }^2 \right) \right.\nonumber \\
&& \left. + 
  q\,\left( 1 + q + \left( -3 + \left( -3 + q \right) \,q \right) \,\lambda  + 
     \left( 1 + q \right) \,{\lambda }^2 \right) \right)\\
h_{\twover}[U_p]&=&\frac{\lambda^{p-1}}{{\left( -1 + q \right) }^2\,\left( 1 + q \right) }
\left( -2\,q^{1 - p}\,\left( q - \lambda  \right) \,\left( -1 + \lambda  \right)  + 
  {\left( -1 \right) }^p\,\left( -1 + q \right) \,q\,\left( -1 + {\lambda }^2 \right) \right.\nonumber \\
&& \left. - 
  \left( 1 + q \right) \,\left( \lambda  + q\,
      \left( 1 + \lambda \,\left( -4 + q + \lambda  \right)  \right)  \right) \right)
\end{eqnarray}
These results alongwith eqs.(\ref{reh}) and (\ref{hath}) give the
the integer BPS invariants. For unknot with framing $p=2$, the integers
BPS invariants are
\begin{center}
\begin{tabular}{r|rr} \hline
g & Q=1/2 & 3/2 \\ \hline
0 & -2 & 2 \\ \hline
\end{tabular}
\hspace{0.5in}
\begin{tabular}{r|rr} \hline
g & Q=2 & 3 \\ \hline
0 & 3 & -2 \\ \hline
\end{tabular}
\hspace{0.5in}
\begin{tabular}{r|rrr} \hline
g & Q=1 & 2 & 3 \\ \hline
0 & -2 & 7 & -4 \\
1 & 0 & 2 & -2 \\ \hline
\end{tabular}

$N^{c=0}_{\one,g,Q}$ \hspace{1in}
$N^{c=0}_{\twohor,g,Q}$ \hspace{1in}
$N^{c=0}_{\twover,g,Q}$
\end{center}

\subsection{Computation of $SO(N)$ reformulated invariants 
$g_{R_1,R_2, \ldots R_r}(q,\lambda)$}
We have computed the functions $g_{R_1, \ldots
R_r}(q,\lambda)$ for framed unknot $U_p$, some framed torus knots 
(${\cal K}_3^{(p)}$, ${\cal K}_5^{(p)}$), framed Hopf link 
($H^{(p_1,p_2)}$) and the connected sum of two framed 
knots ${\cal K}_1 \# {\cal K}_2$. All these invariants
obey the conjectured form (\ref{horg}). 
For completeness, we present $g_R$ invariant for framed unknot
$U_p$ for few representations:
\begin{eqnarray}
g_{\threehor} &=& \frac{\lambda^{3p/2-1}}{{\left( -1 + q \right) }^2\,\left( 1 + q \right) }
\left( {\left( -1 \right) }^p\,q\,\left( -1 + q^p \right) \,\left( -1 + \lambda  \right) \right.\nonumber \\
&& \left.  \left( 1 + q - q^p - q^{2\,p}  - \lambda  - q\,\lambda  + q^{1 + p}\,\lambda  + q^{1 + 2\,p}\,\lambda  \right) \right)\\
g_{\mixed} &=& \frac{-\lambda^{3p/2-1}}{{\left( -1 + q \right) }^2\,\left( 1 + q \right) }
{\left( -1 \right) }^p\,q^{1 - p}\,\left( -1 + q^p \right) \,\left( -1 + \lambda  \right) \,
  \left( 1 - q\,\left( -2 + \lambda  \right)\right.\nonumber \\
&& \left.  + q^p\,\left( -2 + \lambda  \right)  - 2\,\lambda  + 
    q^{1 + p}\,\left( -1 + 2\,\lambda  \right)  \right)\\
g_{\threever} &=& \frac{\lambda^{3p/2-1}}{{\left( -1 + q \right) }^2\,\left( 1 + q \right) }
{\left( -1 \right) }^p\,q^{\frac{1}{2} - 3\,p}\,\left( -1 + \lambda  \right) \,
  \left( -q^{\frac{3}{2}} + q^{\frac{1}{2} + 2\,p}\,\left( 1 - 2\,\lambda  \right) \right.\nonumber \\
&& \left. - 
    q^{\frac{3}{2} + 2\,p}\,\left( -2 + \lambda  \right)  + 
    q^{\frac{1}{2} + 3\,p}\,\left( -1 + \lambda  \right)  + 
    q^{\frac{3}{2} + 3\,p}\,\left( -1 + \lambda  \right)  + {\sqrt{q}}\,\lambda  \right)
\end{eqnarray}
Substituting values for $p$, the above equations reduce to the conjectured
form (\ref {horg}).
\subsection{$N_{(R_1, \ldots R_r),g,Q}^{c=1}$ and 
$N_{(R_1, \ldots R_r),g,Q}^{c=2}$ Computation}
We have computed the integer coefficients corresponding
to cross-cap $c=1$ and $c=2$ unoriented open string amplitude
for various framed knots and framed links using eqns.(\ref {reg}, \ref {hatg}).
The $c=1$ BPS integers exactly matches our earlier paper results\cite{prav}.
Both $c=1$ and $c=2$ integer BPS coefficients for torus knots with 
$p=0$ framing agrees with the results in Refs.\cite{mar9,stev}.
We present the $c=2$ BPS integer coefficients for few framed knots: 

\noindent
\begin{tabular}{r|rrrr} \hline
g & Q= 4 & 5 & 6 & 7 \\ \hline
0 & 21 & -63 & 63 & -21 \\
1 & 70 & -231 & 231 & -70 \\
2 & 84 & -322 & 322 & -84 \\
3 & 45 & -219 & 219 & -45 \\
4 & 11 & -78 & 79 & -11 \\
5 & 1 &-14 & 14 & -1 \\
6 & 0 & -1 & 1 & 0 \\ \hline
\end{tabular} \hspace{.5in}
\begin{tabular}{r|rrrr} \hline
g & Q= 4 & 5 & 6 & 7 \\ \hline
0 & 28 & -84 & 84 & -28 \\
1 & 126 & -406 & 406 & -126 \\
2 & 210 & -756 & 756 & -210 \\
3 & 165 & -705 & 705 & -165 \\
4 & 66 & -363 & 363 & -66 \\
5 & 13 & -105 & 105 & -13 \\
6 & 1 & -16 & 16 & -1 \\
7 & 0 & -1 & 1 & 0 \\ \hline
\end{tabular}

$N^{c=2}_{\twohor,g,Q}$ for knot ${\cal K}_3^{(1)}$ \hspace{1.2in}
$N^{c=2}_{\twover,g,Q}$ for knot ${\cal K}_3^{(1)}$
\vskip.7cm

%\underline{Torus knot $(2,5)$ with framing $p=1$}
{\small
\begin{tabular}{r|rrrr} \hline
g & Q=6 & 8 & 9 & 11 \\ \hline
0 & 55 & -275 & 275 & -55 \\
1 & 495 & -2750 & 2750 & -495 \\
2 & 1716 & -11110 & 11110 & -1716 \\
3 & 3003 & -24090 & 24090 & -3003 \\
4 & 3003 & -31746 & 31746 & -3003 \\
5 & 1820 & -27118 & 27118 & -1820 \\
6 & 680 & -15503 & 15503 & -680 \\
7 & 153 & -5985 & 5985 & -153 \\
8 & 19 & -1540 & 1540 & -19 \\
9 & 1 & -253 & 253 & -1 \\
10 & 0 & -24 & 24 & 0 \\
11 & 0 & -1 & 1 & 0 \\ \hline
\end{tabular} \hspace{.5in}
\begin{tabular}{r|rrrr} \hline
g & Q=6 & 8 & 9 & 11 \\ \hline
0 & 66 & -330 & 330 & -66 \\
1 & 715 & -3905 & 3905 & -715 \\
2 & 3003 & -18656 & 18656 & -3003 \\
3 & 6435 & -47905 & 47905 & -6435 \\
4 & 8008 & -75218 & 75218 & -8008 \\
5 & 6188 & -77415 & 77415 & -6188 \\
6 & 3060 & -54248 & 54248 & -3060 \\
7 & 969 & -26333 & 26333 & -969 \\
8 & 190 & -8855 & 8855 & -190 \\
9 & 21 & -2024 & 2024 & -21 \\
10 & 1 & -300 & 300 & -1 \\
11 & 0 & -26 & 26 & 0 \\
12 & 0 & -1 & 1 & 0 \\ \hline
\end{tabular}
}

$N^{c=2}_{\twohor,g,Q}$ for knot ${\cal K}_5^{(1)}$ \hspace{2in}
$N^{c=2}_{\twover,g,Q}$ for knot ${\cal K}_5^{(1)}$

\vskip.7cm
\noindent
For the connected sum of trefoil with trefoil (${\cal K}_3^{(0)}\#
{\cal K}_3^{(0)}$), we have computed the BPS integers. For
$\twohor$ representation, the integers 
$N^{c=2}_{\twohor,g,Q}$ are 
\begin{center}
{\small
\begin{tabular}{r|rrrrrrrr} \hline
g & Q=4 & 5 & 6 & 7 & 8 & 9 & 10 & 11 \\ \hline
0 & -46 & 627 & -2210 & 3524 & -2891 & 1190 & -193 & -1 \\
1 & -115 & 2857 & -12709 & 23835 & -22244 & 10068 & -1517 & -121 \\
2 & -114 & 5764 & -32974 & 73721 & -78050 & 37511 & -4648 & -1210 \\
3 & -54 & 6412 & -48952 & 133320 & -160443 & 79607 & -5171 & -4719 \\
4 & -12 & 4241 & -45575 & 155369 & -214257 & 107758 & 1914 & -9438 \\
5 & -1 & 1707 & -27770 & 122272 & -195972 & 98868 & 11907 & -11011 \\
6 & 0 & 410 & -11234 & 66279 & -126105 & 63513 & 15145 & -8008 \\
7 & 0 & 54 & -2987 & 24753 & -57626 & 28938 & 10608 & -3740 \\
8 & 0 & 3 & -501 & 6247 & -18593 & 9314 & 4652 & -1122 \\
9 & 0 & 0 & -48 & 1016 & -4138 & 2070 & 1309 & -209 \\
10 & 0 & 0 & -2 & 96 & -604 & 302 & 230 & -22 \\
11 & 0 & 0 & 0 & 4 & -52 & 26 & 23 & -1 \\
12 & 0 & 0 & 0 & 0 & -2 & 1 & 1 & 0 \\ \hline
\end{tabular}
}

%$N^{c=2}_{\twohor,g,Q}$ for ${\cal K}_3^{(0)}\# {\cal K}_3^{(0)}$
\end{center}
\section{Summary and Discussions}
We have explicitly demonstrated the direct evaluation of invariants of
some framed knots and links  carrying composite
representations in $U(N)$ Chern-Simons gauge theory.
Particularly, we proposed a specific choice for the $U(1)$ charge
corresponding to the composite representations (\ref {framchar}) so that 
the composite invariants for framed knots and links are
polynomials in variables $q,\lambda$. Further, this direct method 
enabled us to verify generalised Rudolph's theorem for many 
framed knots (\ref {rud}).

\noindent
The composite invariants was very essential to obtain the untwisted
sector open topological string amplitude(\ref {con1a}) 
on the orientifold of the resolved conifold geometry. 
Similar to Ooguri-Vafa conjecture (\ref {unrfm}),
Marino\cite {mar9} conjectured a form for the
reformulated invariants (\ref {horg}) and the 
refined reformulated invariants (\ref {hath}).
We have verified the conjecture for many framed knots and links
and presented the reformulated invariants for few examples.
The cross-cap $c=0$ BPS integer coefficients (\ref {hath}) 
are also tabulated for these examples.

In earlier works\cite {vinc1,vinc2,prav}, there was  difficulty 
in seperating $c=0$ and $c=2$
contribution from  the topological string free energy(\ref {orie1}) 
but using the parity argument in variable $\sqrt{\lambda}$, 
the cross-cap $c=1$ amplitude could be
determined. With the present work on composite
invariants following the approach \cite {mar9},
we can determine the unoriented topological string amplitude (\ref {orie1}) 
by subtracting the untwisted sector
contribution from the free energy of the open topological string theory on
the orientifold. We have checked that the reformulated
$SO$ invariants obtained from the unoriented topological string free
energy also obeys Marino's conjectured
form (\ref {horg}). Further, the refined $SO$ reformulated invariants
obtained using eqn. (\ref {reg}) satisfies the conjectured form(\ref {hatg}).
We have tabulated the BPS integer invariants corresponding to 
cross-cap $c=2$ obtained from reformulated invariants (\ref {hatg}) for some 
framed knots. The $c=1$ integer coefficients agrees
with our earlier work \cite {prav}. The BPS integer coefficients for 
the standard framing ($p=0$) torus knots and 
torus links agrees with the results in Ref.\cite {stev}.
The verification of Marino's conjectures for many framed knots and 
two-component framed links indirectly confirms that our choice of the 
$U(1)$ charge (\ref {framchar}) for the composite representations 
is correct.

The Marino's conjectures, which we verified for some torus knots and 
torus links, should be obeyed by non-torus knots and non-torus links 
as well. The Chern-Simons approach requires the 
$SU(N)$ quantum Racah coefficients for the non-torus knot 
invariant evaluation. Unfortunately,
these coefficients are not available in the literature.
In Ref.\cite {rama}, the $SU(N)$ quantum Racah coefficients for
some representations could be determined using isotopy 
equivalence of knots enabling evaluation of non-torus knot 
invariants. We believe that there must be a similar
approach of determining composite invariants for the non-torus
knots.

It will be interesting to generalise these integrality properties
in the context of Khovanov homology \cite {gukov} and
Kauffman homology \cite {gukov1}. We hope
to report on this work in a future publication.
\newpage

\begin{center}
{\bf Acknowledgments}
\end{center}
CP would like to express her deep sense of gratitude to 
Prof. Raghava Varma for giving encouragement and providing all the 
necessary help during the course of this work.She is also very much thankful 
to Dr.Ankhi Roy for her sincere cooperation which helped her to 
complete this work. PR would like to thank IRCC, IIT Bombay
for the research grant. CP would like to thank
the organisers for giving her the opportunity
to present this work in  the national strings meeting 
(NSM) held at IIT Bombay during Feb 10-15, 2010. 
\vskip.5cm

\appendix {{\Large {\bf {Appendix}}} 
\section{$U(N)$ Composite Knot Invariants}
$\bullet$ For the composite representation $\rho_{02} \equiv (\twohor,\one)$
whose highest weight is $\Lambda^{(N-1)}+2\Lambda^{(1)}$, the 
the highest weights and the braiding eigenvalues corresponding to the 
irreducible representations
$R_t \in \rho_{02}\otimes\rho_{02}$ are  
{\small
\begin{displaymath}
\begin{array}{|l|l|l|}\hline
R_t&{\rm highest~ weight}&\lambda_t\\ \hline
R_1 & 2\Lambda^{(N-1)}+4\Lambda^{(1)} &
q^{-3/2} \lambda^{3/2} \\
R_2 & \Lambda^{(N)}+{\Lambda^{(N-2)}}+4\Lambda^{(1)}&
- q^{-1/2} \lambda^{3/2}  \\
R_3 & 2\Lambda^{(N-1)}+{\Lambda^{(2)}}+2\Lambda^{(1)}&
-q^{1/2} \lambda^{3/2} \\
R_4 & \Lambda^{(N)}+{\Lambda^{(N-2)}}+\Lambda^{(2)}+2\Lambda^{(1)}&
q^{3/2} \lambda^{3/2} \\
R_5 & \Lambda^{(N)}+\Lambda^{(N-1)}+3\Lambda^{(1)} & 
q^{1/2} \lambda^2 \\
R_6 &\Lambda^{(N)}+\Lambda^{(N-1)}+\Lambda^{(2)}+{\Lambda^{(1)}}&
q^2 \lambda^2  \\
\hline
\end{array} 
\begin{array}{|l|l|l|}\hline
R_t&{\rm highest~ weight}&\lambda_t\\ \hline
R_7 & 2\Lambda^{(N-1)}+2\Lambda^{(2)}& 
q^{3/2} \lambda^{3/2}  \\
R_8 & \Lambda^{(N)}+{\Lambda^{(N-2)}}+2\Lambda^{(2)}&
-q^{5/2} \lambda^{3/2} \\
R_9& 2\Lambda^{(N)}+2\Lambda^{(1)} 
& q^{3/2} \lambda^{5/2} \lambda \\
R_{10} & 2\Lambda^{(N)}+\Lambda^{(2)}&
-q^{5/2} \lambda^{5/2}  \\
R_{11} & \Lambda^{(N)}+\Lambda^{(N-1)}+3\Lambda^{(1)} &
-q^{1/2}\lambda^2 \\
R_{12} & \Lambda^{(N)}+\Lambda^{(N-1)}+\Lambda^{(2)}+{\Lambda^{(1)}}&
-q^2 \lambda^2 \\ 
\hline
\end{array}
\end{displaymath}
}
Using the above table, we can evaluate directly the composite invariants
for framed torus knots and links obtained as closure of two strand braids 
(\ref {knotcom},\ref {link}). 

\noindent
$\bullet$ For the composite representation $\rho_{03} \equiv (\twover,\one)$ 
whose highest weight is ${\Lambda^{(N-1)}}+{\Lambda^{(2)}}~,$, 
the representations $R_t$ obtained from $\rho_{03}\otimes\rho_{03}$
and the signs of the braiding eigenvalues: 
$\epsilon_{R_t}$ are 
{\small
\begin{displaymath}
\begin{array}{ll}
R_1 = 2\Lambda^{(N-1)}+2\Lambda^{(2)}; \epsilon_1= 1 &
R_2 = \Lambda^{(N)}+{\Lambda^{(N-2)}}+2\Lambda^{(2)};\epsilon_2= -1\\
R_3 = \Lambda^{(N)}+\Lambda^{(N-1)}+{\Lambda^{(2)}}+\Lambda^{(1)};
\epsilon_3 =1 \qquad&
R_4 = 2\Lambda^{(N)}+2\Lambda^{(1)};\epsilon_4 =1\\
R_5 = 2\Lambda^{(N-1)}+\Lambda^{(3)}+\Lambda^{(1)};\epsilon_5 =-1 &
R_6 = \Lambda^{(N)}+\Lambda^{(N-2)}+\Lambda^{(3)}+{\Lambda^{(1)}};\epsilon_6 =1\\
R_7 = 2\Lambda^{(N)}+\Lambda^{(2)}; \epsilon_7 = -1&
R_8 = \Lambda^{(N)}+{\Lambda^{(N-1)}}+\Lambda^{(3)};\epsilon_8 =1\\
R_9 = \Lambda^{(N-1)}+\Lambda^{(N-3)}+\Lambda^{(4)};\epsilon_9 =1 &
R_{10} = \Lambda^{(N-1)}+\Lambda^{(N-3)}+\Lambda^{(4)}; \epsilon_{10} =-1\\
R_{11} = \Lambda^{(N)}+\Lambda^{(N-1)}+\Lambda^{(2)}+\Lambda^{(1)}; \epsilon_{11} = -1&
R_{12} = \Lambda^{(N)}+\Lambda^{(N-1)}+\Lambda^{(3)};\epsilon_{12} = -1
\end{array}
\end{displaymath}
}
For the above irreducible representations, quadratic casimir and
eigenvalues can be computed using eqns.(\ref{casi},\ref{egval}).
With this data, the polynomials of the framed knots ${\cal H}_{\twover,\one}[
{\cal K}]$ and framed links carrying the composite representation 
$(\twover,\one)$ can be computed. 

\noindent
$\bullet$ The invariants of knots carrying the composite representation 
$\rho_{04} \equiv (\twohor,\twohor)$ with 
highest weight $2{\Lambda^{(N-1)}}+2{\Lambda^{(1)}}$ will be useful 
for verifying generalised Rudolph theorem (\ref {rud}). 
The highest weights of the ireducible representations $R_t$
obtained from $\rho_{04}\otimes\rho_{04}$ 
and the signs of the braiding eigenvalues $\epsilon_{R_t}$ are
\vskip.01cm
\noindent
{\small
\begin{tabular}{|l|l|l||l|l|l|}\hline
$R_t$&Highest weight&$\epsilon_{R_t}$&$R_t$&Highest weight&$\epsilon_{R_t}$\\
\hline
$R_1$&$4\Lambda^{(N-1)}+4\Lambda^{(1)}$&1& $R_2$&$\Lambda^{(N)}
+2\Lambda^{(N-1)}+{\Lambda^{(N-2)}}+4\Lambda^{(1)}$&-1 \\
$R_3$&$2{\Lambda^{(N)}}+2{\Lambda^{(N-2)}}+4{\Lambda^{(1)}}$&1 &
$R_4$&${\Lambda^{(N)}}+3{\Lambda^{(N-1)}}+3{\Lambda^{(1)}}$& 1\\
$R_5$& $2{\Lambda^{(N)}}+{\Lambda^{(N-1)}}+{\Lambda^{(N-2)}}
+3{\Lambda^{(1)}}$& 1&$R_6$&${\Lambda^{(N)}}+2{\Lambda^{(N-1)}}+
{\Lambda^{(N-2)}}+ {\Lambda^{(2)}} + 2{\Lambda^{(1)}}$&1\\
$R_7$&$ 2{\Lambda^{(N)}}+2{\Lambda^{(N-2)}}+{\Lambda^{(2)}}+
2{\Lambda^{(1)}}$&-1&
$R_8$&$2{\Lambda^{(N)}}+2{\Lambda^{(N-1)}}+2{\Lambda^{(1)}}$&1\\
$R_9$& $2{\Lambda^{(N)}}+{\Lambda^{(N-1)}}+{\Lambda^{(N-2)}}+
{\Lambda^{(2)}}+{\Lambda^{(1)}}$&-1&
$R_{10}$&$2{\Lambda^{(N)}}+2{\Lambda^{(N-2)}}+2{\Lambda^{(2)}}$&1\\
$R_{11}$& $2{\Lambda^{(N)}}+2{\Lambda^{(N-1)}}+{\Lambda^{(2)}}$& -1&
$R_{12}$&$3{\Lambda^{(N)}}+{\Lambda^{(N-2)}}+{\Lambda^{(2)}}$& 1\\
$R_{13}$&$3{\Lambda^{(N)}}+{\Lambda^{(N-1)}}+{\Lambda^{(1)}}$& 1&
$R_{14}$&$4{\Lambda^{(N)}}$& 1\\
$R_{15}$&$4{\Lambda^{(N-1)}}+{\Lambda^{(2)}}+2{\Lambda^{(1)}}$& -1&
$R_{16}$& $4{\Lambda^{(N-1)}}+2{\Lambda^{(2)}}$& 1\\
$R_{17}$& ${\Lambda^{(N)}}+3{\Lambda^{(N-1)}}+3{\Lambda^{(1)}}$&-1&
$R_{18}$&$2{\Lambda^{(N)}}+{\Lambda^{(N-1)}}+{\Lambda^{(N-2)}}+
3{\Lambda^{(1)}}$& -1\\
$R_{19}$&${\Lambda^{(N)}}+3{\Lambda^{(N-1)}}+{\Lambda^{(2)}}+
{\Lambda^{(1)}}$& 1&
$R_{20}$ &${\Lambda^{(N)}}+3{\Lambda^{(N-1)}}+{\Lambda^{(2)}}+
{\Lambda^{(1)}}$&-1\\
$R_{21}$&${\Lambda^{(N)}}+2{\Lambda^{(N-1)}}+{\Lambda^{(N-2)}}+
2{\Lambda^{(2)}}$& -1&
$R_{22}$&$2{\Lambda^{(N)}}+2{\Lambda^{(N-1)}}+2{\Lambda^{(1)}}$& 1\\
$R_{23}$&$ 2{\Lambda^{(N)}}+2{\Lambda^{(N-1)}}+2{\Lambda^{(1)}}$& -1&
$R_{24}$&$ 2{\Lambda^{(N)}}+{\Lambda^{(N-1)}}+{\Lambda^{(N-2)}}+
{\Lambda^{(2)}}+{\Lambda^{(1)}}$& 1\\
$R_{25}$&$3{\Lambda^{(N)}}+{\Lambda^{(N-2)}}+2{\Lambda^{(1)}}$& -1&
$R_{26}$&$ 3{\Lambda^{(N)}}+{\Lambda^{(N-1)}}+{\Lambda^{(1)}}$& -1
\\ \hline
\end{tabular}
}

\noindent $\bullet$For another composite representation 
$\rho_{05}\equiv (\twover,\twover)$ whose 
highest weight is ${\Lambda^{(N-2)}}+{\Lambda^{(2)}}$,
the irreducible representations $R_t$ 
obtained from $\rho_{05}\otimes\rho_{05}$ and the signs of the 
braiding eigenvalues $\epsilon_t$ are
{\small
\begin{displaymath}
\begin{array}{ll}
R_1 = {\Lambda^{(N-2)}}+{\Lambda^{(N-2)}}+{\Lambda^{(2)}}+{\Lambda^{(2)}};\epsilon_{1} = 1\qquad&
R_2 = {\Lambda^{(N-1)}}+{\Lambda^{(N-3)}}+{\Lambda^{(2)}}+{\Lambda^{(2)}};\epsilon_{2} = -1\\
R_3 = {\Lambda^{(N)}}+{\Lambda^{(N-4)}}+{\Lambda^{(2)}}+{\Lambda^{(2)}};\epsilon_{3} = 1&
R_4 = {\Lambda^{(N-1)}}+{\Lambda^{(N-2)}}+{\Lambda^{(2)}}+{\Lambda^{(1)}};\epsilon_{4} = 1\\
R_5 = {\Lambda^{(N)}}+{\Lambda^{(N-3)}}+{\Lambda^{(2)}}+{\Lambda^{(1)}};\epsilon_{5} = 1&
R_6 = {\Lambda^{(N)}}+{\Lambda^{(N-2)}}+{\Lambda^{(1)}}+{\Lambda^{(1)}};\epsilon_{6} = -1\\
R_7 = {\Lambda^{(N-2)}}+{\Lambda^{(N-2)}}+{\Lambda^{(3)}}+{\Lambda^{(1)}};\epsilon_{7} = -1&
R_8 = {\Lambda^{(N-1)}}+{\Lambda^{(N-3)}}+{\Lambda^{(3)}}+{\Lambda^{(1)}};\epsilon_{8} = 1\\
R_9 = {\Lambda^{(N)}}+{\Lambda^{(N-4)}}+{\Lambda^{(3)}}+{\Lambda^{(1)}};\epsilon_{9} = -1&
R_{10} = {\Lambda^{(N-1)}}+{\Lambda^{(N-1)}}+{\Lambda^{(2)}};
\epsilon_{10} = - 1\\
R_{11} = {\Lambda^{(N)}}+{\Lambda^{(N-2)}}+{\Lambda^{(2)}};\epsilon_{11} = 1&
R_{12} = {\Lambda^{(N-1)}}+{\Lambda^{(N-1)}}+{\Lambda^{(1)}}+{\Lambda^{(1)}};\epsilon_{12} = 1\\
R_{13} = {\Lambda^{(N-1)}}+{\Lambda^{(N-2)}}+{\Lambda^{(3)}};\epsilon_{13} = 1&
R_{14} = {\Lambda^{(N)}}+{\Lambda^{(N-3)}}+{\Lambda^{(3)}};\epsilon_{14} = 1\\
R_{15} = {\Lambda^{(N-2)}}+{\Lambda^{(N-2)}}+{\Lambda^{(4)}};\epsilon_{15} = 1&
R_{16} = {\Lambda^{(N-1)}}+{\Lambda^{(N-3)}}+{\Lambda^{(4)}};\epsilon_{16} = -1\\
R_{17} = {\Lambda^{(N)}}+{\Lambda^{(N-4)}}+{\Lambda^{(4)}};\epsilon_{17} = 1&
R_{18} = {\Lambda^{(N)}}+{\Lambda^{(N-1)}}+{\Lambda^{(1)}};\epsilon_{18} = 1\\
R_{19} = 2{\Lambda^{(N)}};\epsilon_{19} = 1&
R_{20} = {\Lambda^{(N-1)}}+{\Lambda^{(N-2)}}+{\Lambda^{(2)}}+{\Lambda^{(1)}};\epsilon_{20} = -1\\
R_{21} = {\Lambda^{(N)}}+{\Lambda^{(N-3)}}+{\Lambda^{(2)}}+{\Lambda^{(1)}};\epsilon_{21} = -1&
R_{22} = {\Lambda^{(N)}}+{\Lambda^{(N-2)}}+{\Lambda^{(2)}};\epsilon_{22} = -1\\
R_{23} = {\Lambda^{(N)}}+{\Lambda^{(N-2)}}+{\Lambda^{(2)}};\epsilon_{23} = 1&
R_{24} = {\Lambda^{(N-1)}}+{\Lambda^{(N-2)}}+{\Lambda^{(3)}};\epsilon_{24} = -1\\
R_{25} = {\Lambda^{(N)}}+{\Lambda^{(N-3)}}+{\Lambda^{(3)}};\epsilon_{25} =-1&
R_{26} = {\Lambda^{(N)}}+{\Lambda^{(N-1)}}+{\Lambda^{(1)}};\epsilon_{26} = -1
\end{array}
\end{displaymath}
}
The composite invariants ${\cal H}_{(R,R)}[{\cal K}]$ for $R=\twohor$
and $R=\twover$ and the corresponding $SO(N)$ invariants ${\cal G}_R[{\cal K}]$
given in appendix A of Ref.\cite {prav} satisfy the
generalised Rudolph theorem (\ref {rud}).
%\subsection{Connected sum of trefoil and trefoil with framing $p$}
%The composite knot invariants for the 
%connected sum of two knots ${\cal K}_1$ and
%${\cal K}_2$ will be 
%\begin{equation}
%{\cal H}_{(R,S)}[{\cal K}_1\# {\cal K}_2]={ 
%{\cal H}_{(R,S)}[{\cal K}_1] {\cal H}_{(R,S)}[{\cal K}_2]\over 
%{\cal H}_{(R,S)}[U]}~.
%\end{equation}
%Similarly, the  $SO(N)$ invariants ${\cal G}_R[{\cal K}_1\# {\cal K}_2$ 
%will also be product of $SO(N)$ invariants of the two knots 
%normalised by the $SO(N)$ unknot invariant. From these
%invariants, we can compute the $SO$ reformulated invariants
%for the connected sum of two knots. For the connected sum of trefoil, the
%reformulated polynomial for various representations are: 
 

\begin{thebibliography}{99}
\bibitem{sinha}S. Sinha, C. Vafa, ``SO and Sp Chern-Simons at Large N,''
{\tt hep-th/0012136}.
\bibitem{mar9}M.Marino,``String theory and the Kauffman polynomial,''
{\tt arXiv:hep-th/0904:1088}.
\bibitem{mor}H. Morton and N.D.A. Ryder, ``Relations between Kauffman
and Homfly satellite invariants,'' {\tt arXiv:math.GT/0902.1239}.
\bibitem{malda}Juan M. Maldacena, `` The Large N Limit of Superconformal 
Field Theories and Supergravity,'' Adv.Theor.Math.Phys.{\bf 2}:231-252,1998,
{\tt hep-th/9711200}. 
\bibitem{gv1}R. Gopakumar, C. Vafa, 
``M-Theory and Topological Strings, I,''{\tt hep-th/9809187}.
\bibitem{gv2}R. Gopakumar, C. Vafa, 
``On the Gauge Theory/ Geometry Correspondence,''{\tt hep-th/9811131}.
\bibitem{gv3}R. Gopakumar, C. Vafa, 
``M-Theory and Topological Strings, II,'' hep-th/9812127.
\bibitem{wittencs}E. Witten, ``Chern-Simons Gauge Theory as a String Theory,''
{\tt hep-th/9207094}.
\bibitem{ov} H. Ooguri, C. Vafa, ``Knot Invariants and Topological Strings,''
Nucl. Phys. {\bf B577}, 419, (2000), {\tt hep-th/9912123}.
\bibitem{lm}
J. M. F. Labastida, M. Marino, ``Polynomial Invariants for Torus Knots
and Topological Strings,'' {\tt hep-th/0004196}.
\bibitem{taps}
P. Ramadevi, T. Sarkar, ``On Link Invariants and Topological String
Amplitudes,'' Nucl. Phys. {\bf B 600} (2001) 487.
\bibitem{laba}
J. M. F Labastida, M. Marino, C. Vafa, ``Knots, Links and Branes
at Large N,'' JHEP11 (2000) 007, {\tt hep-th/0010102}
\bibitem{mari1} M. Marino, C. Vafa, `` Framed Knots at Large $N$,'' 
{\tt hep-th/0108064}.
%\bibitem{kaul}R.K. Kaul, P. Ramadevi, `` Three-Manifold Invariants
%from Chern-Simons Field Theory with Arbitrary Semi-Simple Gauge Groups,''
%Commun.Math.Phys.. {\bf 217} (2001) 295, {\tt hep-th/0005096}.
\bibitem{ramprav} Pravina Borhade, P. Ramadevi, Tapobrata Sarkar,
`` U(N) Framed Links, Three-Manifold Invariants, and Topological Strings,''
Nucl.Phys. B678 (2004) 656-681.
\bibitem{thoo} 
G. `t Hooft, ``A Planar Diagram Theory for Strong Interactions,''
Nucl. Phys. {\bf B 72} (1974) 461.
\bibitem{marin}M. Marino, `` Chern-Simons theory, matrix integrals and
perturbative three-manifold invariants,'' Commun.Math.Phys. 253 (2004) 25-49.
\bibitem{dijk} R. Dijkgraaf and C. Vafa, `` Matrix models,
topological strings and supersymmetric gauge theories,'' Nucl. Phys.
{\bf B 644} 3 (2002).
\bibitem{akmv} M. Aganagic, A. Klemm, M. Marino and C. Vafa, 
``Matrix model as a mirror of Chern-Simons theory,'' JHEP 0402 (2004) 010.
\bibitem{vinc1}V. Bouchard, B. Florea, M. Marino,``Counting Higher Genus 
Curves with Crosscaps in Calabi-Yau Orientifolds,'' JHEP 0412(2004) 035, 
{\tt hep-th/0405083}
\bibitem{vinc2}V. Bouchard, B. Florea, M. Marino,
``Topological Open String Amplitudes On Orientifolds,'' JHEP 0502 (2005) 002,
{\tt hep-th/0411227}.
\bibitem{prav}Pravina Borhade and P. Ramadevi, ``SO(N) Reformulated Link 
Invariants from Topological Strings,'' {\tt hep-th/0505008},
Nucl. Phys. {\bf B727}(2005)471-498.
\bibitem{rama}P. Ramadevi, T.R. Govindarajan and R.K. Kaul, ``Three Dimensional Chern-Simons Theory as a Theory of Knots and Links III : Compact Semi-simple Group,''
{\tt hep-th/9212110}, Nucl.Phys. {\bf B402}(1993) 548-566.
\bibitem{Rudo}L. Rudolph,``A congruence between link polynomials,''
Math. Proc. Cambridge Philo. Soc. {\bf 107} (1990),319-327.
\bibitem{stev}Sebastien Stevan, ``Chern-Simons Invariants of Torus Knots 
and Links,'' {\tt arXiv:1003:2861}.
\bibitem{compo} K. Koike, ``On the decomposition of tensor products of
the representations of the classical groups: by means of the universal
characters,'' Adv. Math. {\bf 74} (1989) 57.
\bibitem{mor1} Richard J. Hadji, Hugh R. Morton,``
A basis for the full Homfly skein of the annulus,''
{\tt arXiv:math/0408078}, Math. Proc. Camb. Philos. Soc. {\bf 141}(2006)81-100.
\bibitem{mor2} H. R. Morton,
``Integrality of Homfly (1,1)-tangle invariants,''
{\tt arXiv:math/0606336}, Algebraic and Geometric Topology. {\bf 7}(2007)
227-238.
\bibitem{gukov}Sergei Gukov, Albert Schwarz, Cumrun Vafa , ``Khovanov-Rozansky Homology and Topological Strings,'' {\tt hep-th/0412243} ,
Lett.Math.Phys.{\bf 74} (2005) 53-74.
\bibitem{gukov1}Sergei Gukov, Johannes Walcher ,  ``Matrix Factorizations and Kauffman Homology ,'' {\tt hep-th/0512298} .
%\bibitem{av} M. Aganagic, C. Vafa, ``Mirror symmetry, 
%D-branes and counting holomorphic discs,'' {\tt hep-th/0012041}.
%\bibitem{akv}
%M. Aganagic, A. Klemm, C. Vafa, ``Disk instantons, mirror symmetry and the
%duality web,'' Z. Naturforsch. {\bf A 57} (2002), {\tt hep-th/0105045}.
%\bibitem{gjs}
%S. Govindarajan, T. Jayaraman and T. Sarkar, 
%``Disc instantons in linear sigma models,'' Nucl. Phys. {\bf B646}, (2002) 498,
%{\tt hep-th/0108234}.
%\bibitem{lermyw}
%W. Lerche, P. Mayr and N. Warner, `N = 1 special geometry, mixed Hodge
%variations and toric geometry,'' {\tt hep-th/0208039}.
%\bibitem{lickorish}
%W. B. R Lickorish, ``3-manifolds and the Temperley Lieb Algebra,''
%Math. Ann {\bf 290} (1991), 657; ``Three-manifold invariants from combinatorics
%o.f Jones polynomial,'' Pac. J. Math, {\bf 149} (1991) 337.
%\bibitem{labb} J.M.F. Labastida, M. Marino, `` A New Point of
%View in the Theory of Knot and Link Invariants,'' {\tt Math.QA/0104180},
%J. Knot Theory Ramifications {\bf 11} (2002) 173.
%\bibitem{mina}
%M. Agananic, M. Marino, C.Vafa, ``All loop topological string amplitudes
%from Chern-Simons theory,'' {\tt hep-th/0206164.}
%\bibitem{laba} J.M.F. Labastida, E. Perez, ``A relation between the 
%Kauffman and the HOMEFLY polynomials for torus knots,''
%{\tt q-alg/9507031}.
\end{thebibliography}
\end{document}